\documentclass[prl,twocolumn,groupedaddress,nofootinbib, preprintnumbers]{revtex4}
\usepackage{amsmath,amssymb,amscd}
\usepackage{listings}
\usepackage{dsfont}
\usepackage{slashed}
\usepackage{color}
\usepackage{ulem}

\usepackage[pdftex]{graphicx}
\usepackage{epstopdf}
\usepackage{subfigure}
\usepackage{epsfig}

\usepackage{xcolor}
\usepackage[colorlinks=true,
            linkcolor=blue,
            urlcolor=blue,
            citecolor=green,          
            bookmarks=true,
            bookmarksnumbered=true,
            breaklinks=true,
            pdfpagemode=Fullscreen,
            pdfstartview=FitBH]{hyperref}

\definecolor{gesfpurple}{rgb}{0.47,0.19,0.42}

\definecolor{gesflanse}{rgb}{0.00,0.50,0.50}

\definecolor{gesfblue}{rgb}{0.08,0.42,0.76}

\definecolor{gesfred}{rgb}{1,0,0}

\definecolor{gesfwhite}{rgb}{1,1,1}

\definecolor{gesfblack}{rgb}{0,0,0}

\newcommand{\geqn}[1]{\hypersetup{linkcolor=blue}(\ref{#1})\hypersetup{linkcolor=blue}}
\newcommand{\gfig}[1]{{\hypersetup{linkcolor=violet}Fig.~\ref{#1}\hypersetup{linkcolor=blue}}}

\begin{document}

\preprint{KCL-PH-TH/2018-02, CERN-TH/2018-014, IPMU18-0022}

%\title{LHC Constraints on Dimension-8 Operator Contributions to $gg \to \gamma \gamma$}
\title{Constraining Gluonic Quartic Gauge Coupling Operators with $gg \to \gamma \gamma$}

\vskip 0.5in
\author{John~Ellis}
 
 \affiliation{Theoretical Particle Physics and Cosmology Group, Physics Department, 
King's College London, London WC2R 2LS, UK;\\
National Institute of Chemical Physics \& Biophysics, R\"avala 10, 10143 Tallinn, Estonia;\\
Theoretical Physics Department, CERN, CH-1211 Geneva 23, Switzerland}

\author{Shao-Feng~Ge}

\affiliation{Kavli IPMU, UTIAS, University of Tokyo, Kashiwa, Chiba 277-8583, Japan;\\
        Department of Physics, University of California, Berkeley, CA 94720, USA; \\
        Theoretical Physics Department, Fermi National Accelerator Laboratory, Batavia, IL 60510, USA}

\begin{abstract}
Gluon-gluon to photon-photon scattering $gg \to \gamma\gamma$ offers to the LHC experiments a uniquely powerful probe of dimension-8 operators in the Standard Model Effective Field Theory (SMEFT) that are quadratic in
both the electromagnetic and gluonic field-strength tensors, such as would appear in the Born-Infeld extension of the
Standard Model (SM). We use 13-TeV ATLAS data on the production of isolated photon pairs to set lower
limits on the scales of dimension-8 operators  $M \gtrsim 1$~TeV, and discuss the prospective sensitivities of possible future hadron colliders.

\end{abstract}

\maketitle 

{\it Introduction} -- A model-independent way to constrain possible extensions of the SM with high-scale new
physics that decouples at low energies is provided by the SMEFT~\cite{SMEFT},
which employs a systematic expansion in the effective mass dimensions of the new operators generated
by high-scale physics beyond the SM. Apart from the dimension-5 operators that may contribute to
neutrino masses and oscillations~\cite{Weinberg}, the most prominent operators are those of dimension 6, whose
coefficients scale as $1/\Lambda^2$, where $\Lambda$ represents a generic new-physics scale. There
have been many studies of the constraints on dimension-6 operator coefficients imposed by current and
potential future collider data~\cite{data}. Some attention has also been paid to the experimental constraints on operators of
dimension $d \gtrsim 8$, whose effects at low energies are suppressed by ${\cal O}(E/\Lambda)^{d - 4}$, particularly those
involving four electroweak gauge field strengths, see, e.g.,~\cite{Eboli:2006wa, MultiGaugeBoson}.

We focus here on dimension-8 operators in the SMEFT that are 
quadratic in the  field strength tensors of both the gluon fields of QCD, $G^a_{\mu \nu}: a = 1, ..., 8$
and electroweak gauge fields, either the $W^i_{\mu \nu}: i = 1, 2, 3$ of SU(2) or the $B_{\mu \nu}$ of U(1). 
As we review in more detail below, there are 8 independent such dimension-8 operators, 4 involving pairs of the $W^i_{\mu \nu}$
and 4 involving pairs of $B_{\mu \nu}$. Since the electromagnetic field strength tensor $F_{\mu \nu}$ is a
specific combination of $W^3_{\mu \nu}$ and $B_{\mu \nu}$, the $g g \to \gamma \gamma$
scattering process is sensitive to just 4 combinations of these dimension-8 operators. 

One of these combinations is of particular interest, as it arises in the Born-Infeld (BI) extension of the SM
with the following Lagrangian ${\cal L}_{\rm BISM}$:
\begin{equation}
\hspace{-2mm}
  \beta^2
\left[
  1
- \sqrt{1 + \sum_{\lambda = 1}^{12} \frac {F^\lambda_{\mu \nu} F^{\lambda, \mu \nu}}{2 \beta^2}
- \left( \sum_{\lambda = 1}^{12} \frac {F^\lambda_{\mu \nu} \widetilde F^{\lambda, \mu \nu}}{4 \beta^2} \! \right)^2} \;
\right] ,
\label{BISM}
\end{equation}
where $\beta \equiv M^2$ is the BI nonlinearity scale and
the index $\lambda$ runs over the 12 generators of the SM SU(3)$\times$SU(2)$\times$U(1) gauge
group. Born and Infeld proposed a similar nonlinear extension of QED in 1934~\cite{BornInfeld34}, motivated by a `unitarian' idea that 
there should be an upper limit on the strength of the electromagnetic field. However, this theory remained
largely a curiosity until Fradkin and Tseytlin~\cite{FT85} showed in 1985 that it appears
in models inspired by M-theory, e.g., in which vector fields are coupled to matter particles that are
localized on lower-dimensional `branes'~\cite{Tseytlin}. We note also that it has recently been shown that BI
theories have uniquely soft scattering amplitudes in the infrared limit~\cite{Cheung}.

It was pointed out in~\cite{Ellis:2017edi} that a measurement of light-by-light scattering in heavy-ion collisions
at the LHC by the ATLAS Collaboration~\cite{ATLAS17} imposes a constraint on the BI extension of QED
that is orders of magnitude stronger than that available from previous, lower-energy experiments~\cite{previous},
corresponding, e.g.,  in the context of M-theory to an upper limit on the separation between branes
$\lesssim 1/(100~{\rm GeV})$.
The purpose of this Letter is to show that an ATLAS measurement of 
$gg\to \gamma \gamma$ scattering in proton-proton collisions~\cite{Aaboud:2017yyg} strengthens this limit by almost another order of
magnitude in the context of a BI extension of the SM. This bound penetrates
significantly the parameter space of variants of M-theory with large extra dimensions~\cite{Tseytlin}, and 
we show how future hadron colliders would offer even greater sensitivity. 

{\it Dimension-8 Gluon/Photon Operators} -- Constructing the effective operators that
contribute to $g g \rightarrow \gamma \gamma$ scattering needs two
gluon fields and two photon fields~\cite{dipole}. Since fermions and massive vector bosons are absent in the external
states of this process, the candidate operators involve only these fields,
appearing via the gluon field strength $G^a_{\mu \nu}$ and
the photon field strength $F_{\mu \nu}$, which is a combination of the $B_{\mu \nu}$ of $U(1)_Y$ and
the $W^3_{\mu \nu}$ of $SU(2)_L$ with coefficients $s_W \equiv \sin \theta_W$ and $c_W \equiv \cos \theta_W$,
respectively. The dimension-8 operators relevant to
$gg \rightarrow \gamma \gamma$ scattering require
two gluon field strengths $G^a_{\mu \nu}$ and two electroweak field strengths
$B_{\mu \nu}$ or $W^i_{\mu \nu}$. The two colour
indices $a$ must be contracted, as must the two $SU(2)_L$ indices $i$. Lorentz
invariance allows 4 different ways of contracting the 8 space-time indices.
Thus there are 8 independent gluonic quartic gauge coupling (gQGC) operators,
and the relevant dimension-8 part of the effective Lagrangian may be written as
$\mathcal L_{gT} = \sum^7_{i=0} \frac{1}{16 \beta^2_i} \mathcal O_{gT,i}$ where
\begin{subequations}
\begin{eqnarray}
  \mathcal O_{gT,0}
& = &
       \sum_a  G^a_{\mu \nu}  G^{a, \mu \nu}
\times \sum_i  W^i_{\alpha \beta}  W^{i, \alpha \beta} ,
\\
  \mathcal O_{gT,1}
& = &
       \sum_a  G^a_{\alpha \nu}  G^{a, \mu \beta}
\times \sum_i  W^i_{\mu \beta}  W^{i, \alpha \nu} ,
\\
  \mathcal O_{gT,2}
& = &
       \sum_a  G^a_{\alpha \mu}  G^{a, \mu \beta}
\times \sum_i  W^i_{\nu \beta}  W^{i, \alpha \nu} ,
\\
  \mathcal O_{gT,3}
& = &
       \sum_a  G^a_{\alpha \mu}  G^a_{\beta \nu}
\times \sum_i  W^{i, \mu \beta}  W^{i, \nu \alpha} ,
\\
  \mathcal O_{gT,4}
& = &
       \sum_a  G^a_{\mu \nu}  G^{a, \mu \nu}
\times B_{\alpha \beta} B^{\alpha \beta} ,
\\
  \mathcal O_{gT,5}
& = &
       \sum_a  G^a_{\alpha \nu}  G^{a, \mu \beta}
\times B_{\mu \beta} B^{\alpha \nu} ,
\\
  \mathcal O_{gT,6}
& = &
       \sum_a  G^a_{\alpha \mu}  G^{a, \mu \beta}
\times B_{\nu \beta} B^{\alpha \nu} ,
\\
  \mathcal O_{gT,7}
& = &
       \sum_a  G^a_{\alpha \mu}  G^a_{\beta \nu}
\times B^{\mu \beta}  B^{\nu \alpha} ,
\end{eqnarray}
\label{eq:OgTi}
\end{subequations}
where the scales $\sqrt{\beta_i} \equiv M_i$ represents the scales of the physics
beyond the SM that induces these effective operators.
Operators of form similar to $\mathcal O_{gT,3}$ and $\mathcal O_{gT,7}$ have
not been discussed in the context of electroweak QGCs~\cite{Eboli:2006wa}.
The effective Lagrangian for $gg \rightarrow \gamma \gamma$ scattering may
be written in the form $ \mathcal {\hat L}_{gT} = \sum^3_{i=0} \frac{1}{16  {\hat \beta}^2_i}  \mathcal {\hat O}_{gT,i}$,
where $ \frac{1}{{\hat \beta}^2_i} \equiv \frac{s^2_W}{\beta^2_i} + \frac{c^2_W}{\beta^2_{i+4}} , i = 0, 1, 2, 3$ and
the  $\mathcal {\hat O}_{gT,i}$ have the same forms as the $\mathcal O_{gT,i}, i = 4, 5, 6, 7$ but with $B_{\alpha \beta}$
replaced by $F_{\alpha \beta}$, etc.

The first term in the BI extension \geqn{BISM} of the SM
generates $\mathcal O_{gT,0}$ and $\mathcal O_{gT,4}$, and hence also $\mathcal {\hat O}_{gT,0}$.
The second term in (\ref{BISM}) yields a quartic interaction
with the Lorentz structure $(W_{\mu \nu} \widetilde W^{\mu \nu}) (G_{\alpha \beta} \widetilde G^{\alpha \beta}) =
- 2 (W_{\mu \nu} G^{\mu \nu})^2 + 4 (W_{\mu \nu} G^{\nu \alpha} W_{\alpha \beta} G^{\beta \mu})$, where the 
SU(2) and colour indices have been omitted, and similarly with $W \to B$, generating 
$\mathcal O_{gT,1}, \mathcal O_{gT,5}$ and $\mathcal O_{gT,3}, \mathcal O_{gT,7}$, and hence also 
$\mathcal {\hat O}_{gT,1}, \mathcal {\hat O}_{gT,3}$. One would expect the
coefficients $\frac{1}{\beta^2_0}$ and {$\frac{1}{\beta^2_4}$} to be equal at the common BI scale
{$M_0 = M_4$}, but subject to different renormalization below that scale, {and similarly for $i = 1, 3$ and $5, 7$}. However, since the constraint
we find on the BI scale is not very different from the electroweak scale,  this effect is small and
we neglect it in our analysis. In this approximation, the experimental constraints on $M_0$ that we derive below are
proxies for the corresponding constraints on the BI scale $M = \sqrt{\beta}$.

\begin{figure}[t]
\centering
\includegraphics[width=7.2cm]{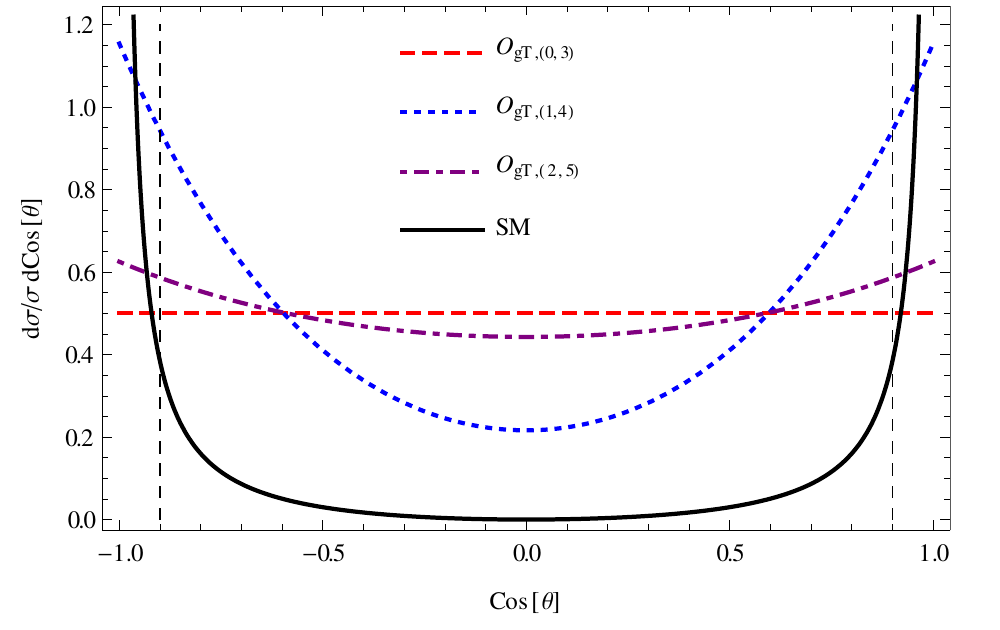}
\centering
\includegraphics[width=7.5cm]{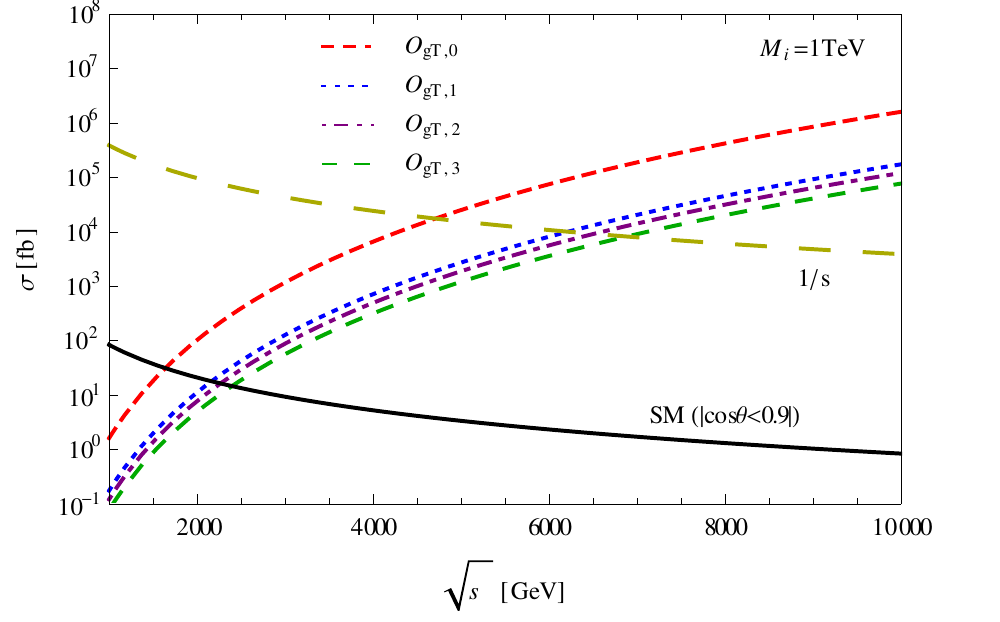}
\caption{Angular distributions in the centre-of-mass frame of $gg \rightarrow \gamma \gamma$ scattering (upper panel) and
the corresponding total cross sections (lower panel), 
where a cut $|\cos \theta| < 0.9$ on the scattering angle
in the center-of-mass frame
has been used to regularize the SM cross section.}
\label{fig:dSigma}
\end{figure}

{\it $gg \rightarrow \gamma \gamma$ Scattering} -- 
The different Lorentz structures in \geqn{eq:OgTi} yield different cross sections for
$gg \rightarrow \gamma \gamma$ process~\cite{PX}:
\begin{equation*}
  \frac {d \sigma_{gT,i}}{d t}
=
  \frac {(s^4_W, c^4_W)}{\beta^4_i}
\begin{cases}
  \frac {s^2}{4096 \pi}  & i = 0, 4 \, , \\[1mm]
  \frac {s^4 - 2 s^2 (t^2 + u^2) + 3 (t^4 + u^4)}{32768 \pi s^2} & i = 1,5 \, , \\[1mm]
  \frac {2 s^4 + t^4 + u^4}{131072 \pi s^2} & i = 2,6 \, , \\[1mm]
  {\frac {s^4 + t^4 + u^4 + 4 t^2 u^2}{131072 \pi s^2}} & i = 3,7 \, ,
\end{cases}
\label{scattering}
\end{equation*}
where $s \equiv (p_{\gamma_1} + p_{\gamma_1})^2$, $t \equiv \frac s 2 (\cos \theta - 1)$ and $u \equiv - t - s$ are Mandelstam variables.
The four different Lorentz structures have different dependences
on the scattering angle $\theta$ in the centre-of-mass frame, independent of $s$,
as shown in the upper panel of \gfig{fig:dSigma}. Once an excess beyond the 
SM background is seen, the Lorentz structures can be identified by fitting
the $\theta$ distribution. These angular dependences can be contrasted
with that of the SM $q \bar q \rightarrow \gamma \gamma$ background:
$d \sigma / d \cos \theta \propto \cot^2 \theta$ that vanishes for $\theta = \pi/2$
 in the massless limit. A cut on the angular distribution $|\cos \theta| \leq 0.8$ or 0.9 (equivalent to a cut in pseudorapidity) would be effective
for suppressing the SM background. 

Each gauge field contributes a momentum factor to the amplitudes generated by the dimension-8 gQGC operators (2), so the total cross sections scale as $s^3$,
\begin{equation}
  \sigma_{gT,i}
=
  \frac {(s^4_W, c^4_W)}{4096 \pi \beta^4_i}
\times
  \left( 1, \frac {13}{120}, \frac 3 {40}, {\frac{23}{480}} \right)
\times
  s^3 \,.
\label{eq:sigma}
\end{equation}
The cross sections for 
{$\mathcal O_{gT,(1,5)}$, $\mathcal O_{gT,(2,6)}$, and $\mathcal O_{gT,(3,7)}$ are
roughly one order smaller than those for $\mathcal O_{gT,(0,4)}$, respectively}.
Considering the $s^4_W$ and $c^4_W$ coefficients, the contributions of the eight gQGC operators have a hierarchical structure:
{$\sigma_{gT,4} \approx 10 \sigma_{gT,(0,5,6,7)} \approx 100 \sigma_{gT,(1,2,3)}$}
for identical scales of $M_i$.

A characteristic energy of $g g \rightarrow \gamma \gamma$ scattering at LHC-13~TeV
is $\mathcal O(1)\,\mbox{TeV}$, which places a natural limit on the applicability
of gQGC operators.
If $\sqrt s$ exceeds $M_i$, the $\mathcal O_{gT,i}$ operators
would eventually cease to be a good approximation, violating the unitarity constraint.
To preserve unitarity, we assume that the cross
section falls with the diphoton invariant mass, $\sigma \approx 1/s = 1/m^2_{\gamma \gamma}$,
above the scales  $\sqrt{s_i}$ where unitarity is saturated:
\begin{equation}
  \sqrt{s_i}
=
  M_i
\left[
  \frac {(s^4_W, c^4_W)}{4096 \pi}
\left( 1, \frac {13}{120}, \frac 3 {40}, {\frac {23}{480}} \right)
\right]^{- \frac 1 8} \,,
\end{equation}
corresponding to ratios
$\sqrt{s_i}/M_i =$ 4.71, 6.21, 6.51, 6.88, 3.49, 4.60,4.82, 5.10 for $i = 0, \cdots, 7$.
The cross section increases as $s^3$ below and decreases as $1/s$ above
the saturation point.

\begin{figure}[t]
\centering
\includegraphics[width=7.5cm]{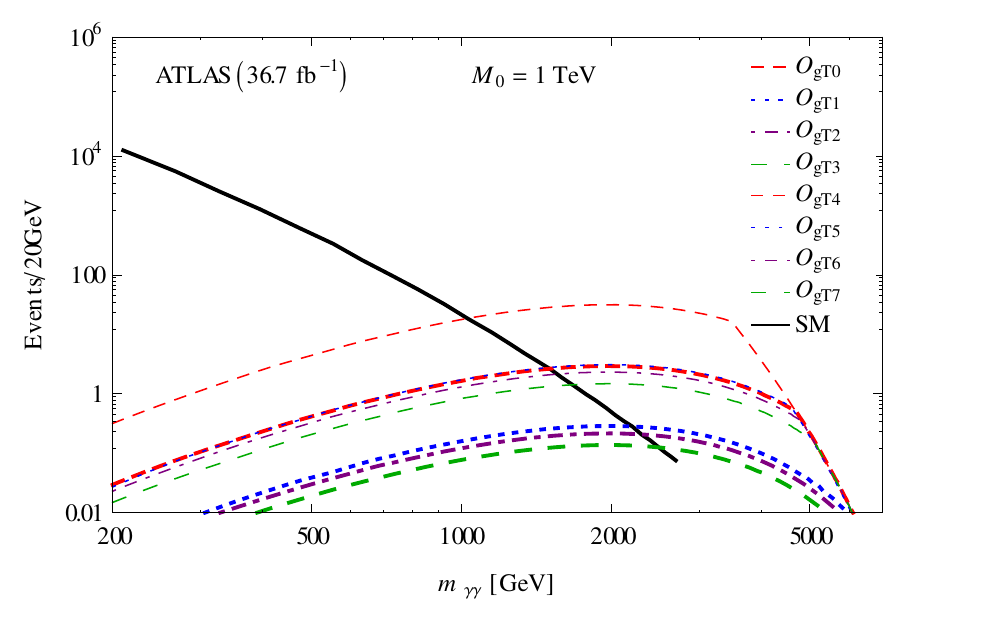}
\includegraphics[width=7.cm]{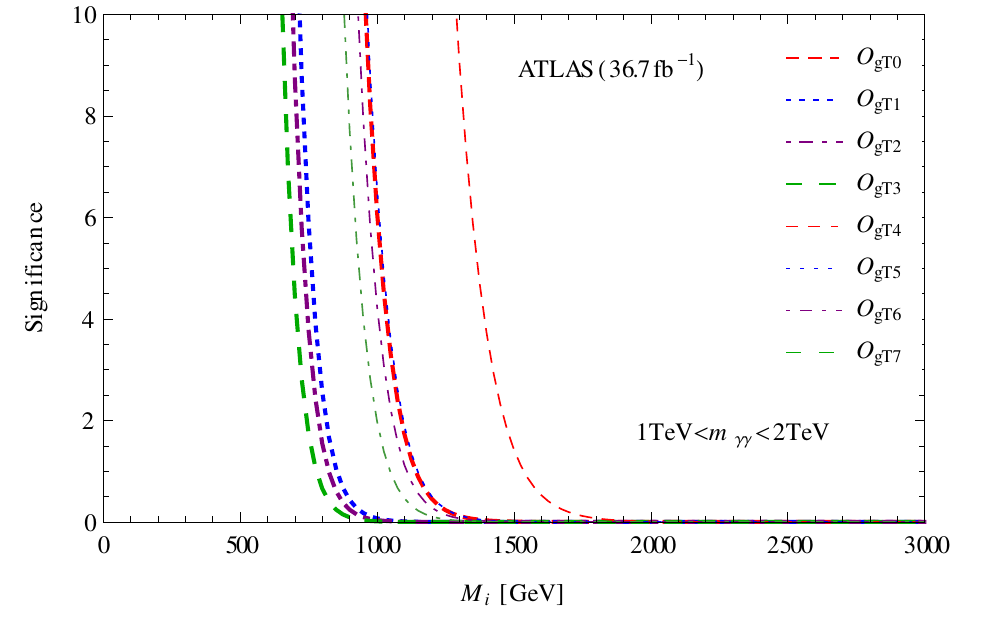}
\caption{Event spectra (upper panel) and sensitivities (lower panel) of ATLAS with $36.7\,\mbox{fb}^{-1}$.}
\label{fig:LHCspectrumsensitivity}
\end{figure}

The cross section for the SM background
$q \bar q \rightarrow \gamma \gamma$ also falls with energy: at leading order:
$\sigma_{\rm SM} \approx \sum_q Q^4_q [ \log (1 / (1 - \cos \theta_{cut})) - 2]/24 \pi s$, where
$Q_q$ is the electric charge and $\theta_{cut}$ is a cut 
$|\cos \theta| < \cos \theta_{cut} = 0.9$ on the  scattering angle in the center-of-mass frame.
Higher-order QCD
corrections increase this by a slowly-varying K-factor \cite{Kfactor},
yielding results in agreement with
with the ATLAS measurements~\cite{Aaboud:2017yyg}.
The lower panel of \gfig{fig:dSigma} compares this background
with the gQGC signals for $M_i = 1$~TeV. The SM background lies
far below the scale of unitarity saturation, and falls below the 
potential gQGC signals when $\sqrt s \sim 2$ to $3$~TeV.
In addition to exploiting the different angular distribution,
one may also suppress the SM background by cutting
low $\sqrt s$ events.

{\it Constraints from the ATLAS Data} -- The ATLAS Collaboration searched for
new physics with high-mass diphoton final states \cite{Aaboud:2017yyg}.
Comparing their searches for spin-0 and -2 resonant and non-resonant
signals, the non-resonant case is the closest in spirit to the
gQGC contribution to $gg \rightarrow \gamma \gamma$ that we consider
here.

The ATLAS analysis uses the fiducial region $|\eta| < 2.37$, excluding
the blind transition region $1.37 < |\eta| < 1.52$.
In the non-resonant signal search, the photon transverse energies
receive a cut $E_T > 55\,\mbox{GeV}$.
For non-resonant Kaluza-Klein signal, the geometric acceptance increases from 58\% at
$M_S = 3.5\,\mbox{TeV}$ to 65\% at $M_S = 5\,\mbox{TeV}$ where $M_S$
is the cutoff scale of the Kaluza-Klein spectrum. 

In our analysis, we assume a constant geometric acceptance of 60\%, which is to be
combined with the efficiency for reconstruction and identification that 
is approximately constant at 77\%, yielding an overall signal event selection
efficiency of 46\%. The upper panel of \gfig{fig:LHCspectrumsensitivity} shows
the expected signal event rates at ATLAS with 36.7~fb$^{-1}$ and 13 TeV for the
different gQGC operators with $M_i = 1\,\mbox{TeV}$, as functions of the
invariant mass $m_{\gamma \gamma}$ \cite{FR,Calc}. 
For comparison, we also show the background
rate extracted from the background-only fit in the Fig.~2b of \cite{Aaboud:2017yyg}.
With cutoff scale $M_i = 1\,\mbox{TeV}$, the background and
gQGC signals cross around $m_{\gamma \gamma} = 1$ to $2\,\mbox{TeV}$
as expected from \gfig{fig:dSigma}. Above $m_{\gamma \gamma} = 1.5\,\mbox{TeV}$,
the signal rate keeps rising before saturating unitarity at $m_{\gamma \gamma} \gtrsim 4\,\mbox{TeV}$,
depending on the model.
In our estimations of the SM background and signals we use the cut
$m_{\gamma \gamma} < 2\,\mbox{TeV}$, well below these unitarity saturation scales.

With this cut, we make a binned analysis of the ATLAS data \cite{Aaboud:2017yyg}
to quantify the sensitivity $\sum_i |S_i + B_i - N_i|/\sqrt{N_i}$
to these operators, where $S_i$ and $B_i$ are the predicted total signal and
background while $N_i$ is the number of events actually measured
in the $i$-th bin. We plot the significances, evaluated using the $\Delta \chi^2$ distributions,
as functions of the nonlinearity scale $M_i$ in the lower panel of \gfig{fig:LHCspectrumsensitivity}.
The significances decrease very rapidly with $M_i$. Since the gQGC
operator coefficients are suppressed by $1/M^4_i$, the cross sections
fall as $1/M^8_i$ and small changes in the $M_i$ can affect the significances
dramatically.
The hierarchical structure of the cross sections generated by the eight gQGC operators is
manifest in the 95\% C.L. lower bounds derived from the ATLAS data:
$M_i \gtrsim (1040, 777, 750, 709, 1399, 1046, 1010, 954)$~GeV~\cite{Plehn}.
We recall that $M_0$ is a proxy for the SM
BI scale $M$. Note that setting a $2\,\mbox{TeV}$ upper cut on invariant mass
$m_{\gamma\gamma}$ is roughly the same as shifting the saturation point
down to $\sqrt{s_i} = 2\,\mbox{TeV}$.

\begin{figure}[t]
\centering
\includegraphics[width=7cm]{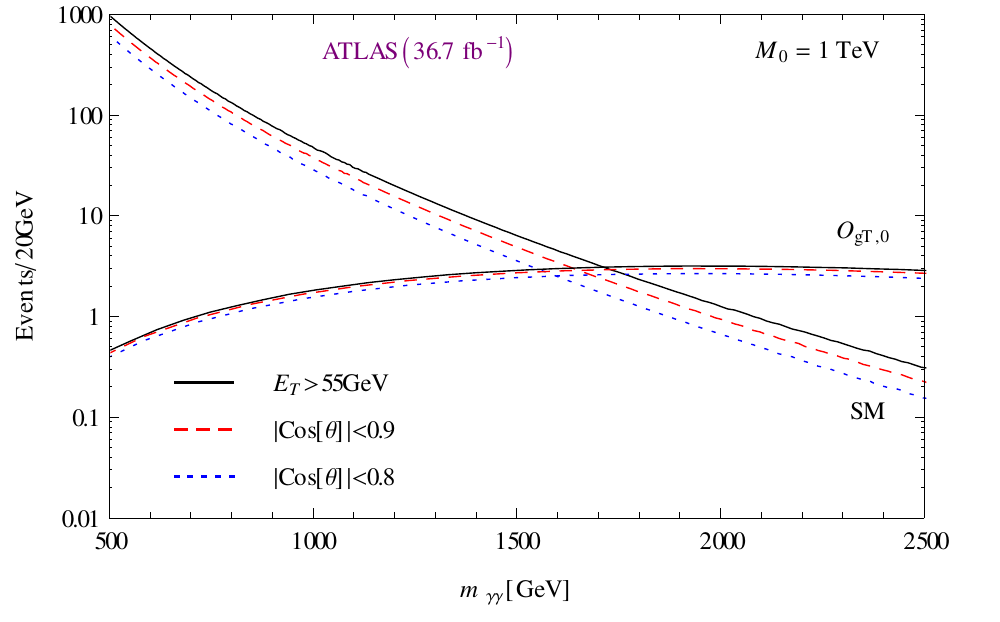}
\caption{The effects of $E_T$ and $\cos \theta$ cuts for the ATLAS search.} 
\label{fig:cuts}
\end{figure}

{\it Sensitivities at Future Hadron Colliders} --
As discussed analytically above, the most effective cut for
suppressing the SM background is that on the scattering angle in the centre-of-mass
frame. \gfig{fig:cuts} shows the cross sections obtained with different
cuts on the angular distributions, applying the fiducial cut $|\eta(\gamma)| < 2.37$
in all cases. In comparison, the black curves
were obtained with the ATLAS $E_T$ cut,
$E_T(\gamma) > 55\,\mbox{GeV}$.
Across the whole invariant-mass region, the scattering angle cut
can reduce significantly the SM background,
with much smaller effects on the signals.
Therefore, the scattering angle cut in the
center-of-mass frame is more suitable than $E_T$ cut for the search
for dimension-8 gQGC operators. In following discussion of the sensitivities
at future colliders, we
apply cuts $|\eta(\gamma)| < 2.37$ and $|\cos \theta(\gamma)| < 0.8$. 

\begin{figure}[t]
\centering
\includegraphics[width=7cm]{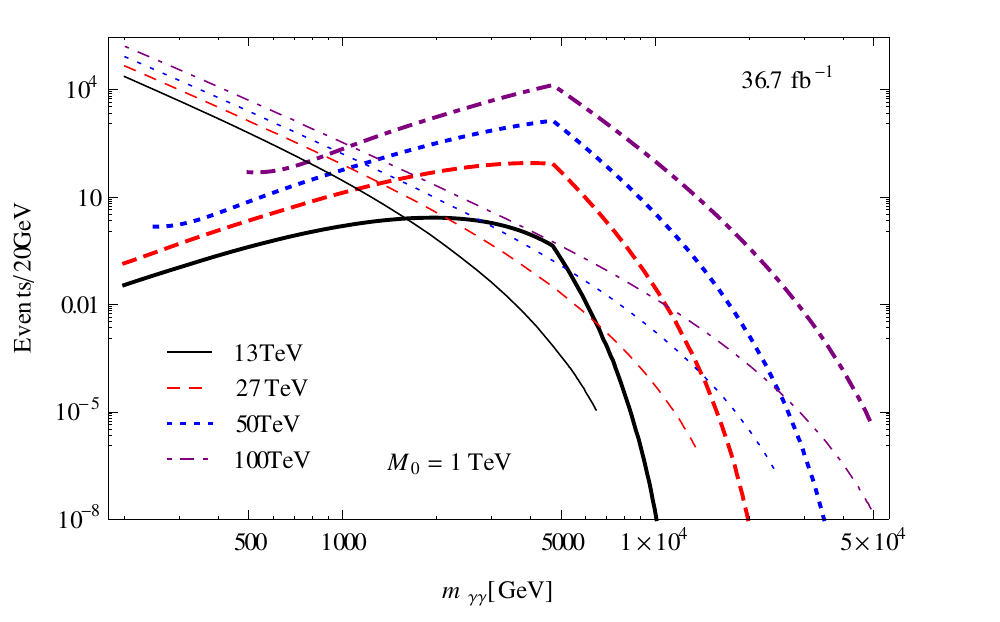}
\includegraphics[width=7.5cm]{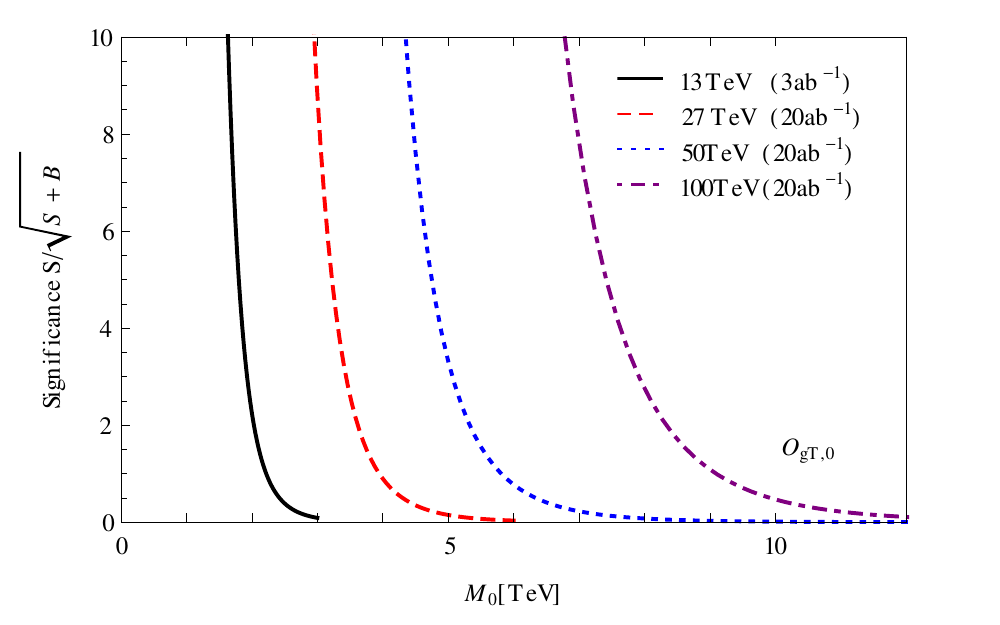}
\caption{The event spectra (upper panel) and sensitivities (lower panel) at future hadron colliders.}
\label{fig:chi2}
\end{figure}

The upper panel of \gfig{fig:chi2} shows how the SM background (thinner lines) and the gQGC signals
(thicker lines) change with collider energy. For illustration, we only plot
$\mathcal O_{gT,0}$ whose features are shared by other gQGC operators.
For comparison,
we use the same luminosity $36.7\mbox{fb}^{-1}$ for different collider
energies, noting that the event rates would be much larger for
the expected luminosities at HE-LHC at 27~TeV~\cite{HE-LHC}, FCC-hh~\cite{FCC-hh} 
and SppC~\cite{SppC}. At higher energies the signal
event spectra are significantly enhanced, especially in the high $m_{\gamma\gamma}$
range. For $\sqrt s = 100\,\mbox{TeV}$, the event rate can be as large
as $10^4/20\,\mbox{GeV}$ {for $M_i = 1\,\mbox{TeV}$}.
Although the SM background also increases, its contribution
at high $m_{\gamma \gamma}$ range is still negligibly small.
At higher collider energies $\sqrt s$, the crossing point between the SM
background and the gQGC signal curves decreases.
The crossings for $\sqrt s = (13, 27, 50, 100) \, \mbox{TeV}$ happen at
$m_{\gamma\gamma} = (1.60, 1.32, 1.17, 1.06) \, \mbox{TeV}$.
We increase the upper cut on the invariant mass roughly
proportionally to the collider energy: $m_{\gamma \gamma} \leq (3,5,9,14)\,\mbox{TeV}$,
respectively.

The lower panel of \gfig{fig:chi2} shows the significances
$\sum_i S_i /\sqrt{S_i + B_i}$
at various hadron colliders, including LHC at 13~TeV with $3\,\mbox{ab}^{-1}$ and
HE-LHC at 27~TeV, FCC-hh and SppC at 50 or 100TeV, each with $20\,\mbox{ab}^{-1}$
for the $\mathcal O_{gT,0}$ operator.
Enhancing collider energy and luminosity significantly improves the
sensitivity. The 3-$\sigma$ discovery sensitivity can reach
$2.1\,\mbox{TeV}$ at 3\,ab$^{-1}$ LHC,
$4.5\,\mbox{TeV}$ at the 27-TeV HE-LHC, $7.5\,\mbox{TeV}$ at the 50-TeV versions and
$13\,\mbox{TeV}$ at the 100-TeV versions of FCC-hh and SppC.
For FCC-hh and SppC at 100TeV, the sensitivity would be another order of magnitude
better than the current ATLAS analysis with $36.7\mbox{fb}^{-1}$ at 13~TeV,
well into the range of potential interest to string models.

{\it Conclusions} --
The ATLAS data on light-by-light scattering in heavy-ion collisions can exclude
the QED BI~\cite{BornInfeld34} scale $\lesssim 100$~GeV\cite{Ellis:2017edi}.
In this paper we have shown that the ATLAS data on $g g \rightarrow \gamma \gamma$
scattering enhances the sensitivity by an order of magnitude, to
$\gtrsim 1$~TeV for the analogous dimension-8 operator scales
containing other combinations of gluon and electromagnetic fields.
{This constraint on the BI extension of SM} is very
interesting in view of its connections with string theory~\cite{FT85}
and particularly models in which branes are separated by distances $\gtrsim 1$~TeV$^{-1}$~\cite{Tseytlin}.
Moreover, similar searches for $\gamma \gamma$ production at possible future
proton-proton colliders could be sensitive to BI scales in the multi-TeV scale,
complementing the searches via dimension-6 SMEFT operators
\cite{dim6}.

\section*{Acknowledgements}

{We thank Sasha Belyaev for useful discussions and Alexander Pukhov
for help with CalcHEP.}
The work of J.E. was supported partly by the STFC Grant ST/L000258/1,
and partly by the Estonian Research Council via a Mobilitas Pluss grant.
The work of S-F.G. was supported by JSPS KAKENHI Grant Number JP18K13536,
World Premier International (WPI) Research Center Initiative, MEXT, Japan,
and Fermilab Neutrino Physics Center Fellowship Program.


\begin{thebibliography}{99}
\bibitem{dipole}
Although the dimension-6 magnetic dipole operators
$\bar Q \gamma^\mu W_{\mu \nu} D^\nu Q + h.c.$ and
$\bar Q \gamma^\mu B_{\mu \nu} D^\nu Q + h.c.$ modify
the $q \bar q A$ vertex, their prospective $t$-channel contribution to 
$q \bar q \rightarrow \gamma \gamma$ vanishes when the photons are
on shell.
\end{thebibliography}
\end{document}